# Origin of the laser-induced picosecond spin current across magnetization compensation in ferrimagnetic GdCo

*Guillermo Nava Antonio, Quentin Remy, Jun-Xiao Lin, Yann Le Guen, Dominik Hamara, Jude Compton-Stewart, Joseph Barker, Thomas Hauet, Michel Hehn, Stéphane Mangin\*, and Chiara Ciccarelli\**

G. Nava Antonio, D. Hamara, C. Ciccarelli
Cavendish Laboratory, University of Cambridge, CB3 0HE Cambridge, UK
E-mail: cc538@cam.ac.uk

Q. Remy
Department of Physics, Freie Universität Berlin, 14195 Berlin, Germany

J.-X. Lin, Y. Le Guen, J. Compton-Stewart, T. Hauet, M. Hehn, S. Mangin
Institut Jean Lamour, CNRS UMR 7198, Université de Lorraine, F-54506 Nancy, France
E-mail: stephane.mangin@univ-lorraine.fr

J. Barker
[4]School of Physics and Astronomy, University of Leeds, LS2 9JT Leeds, UK

The optical manipulation of magnetism enabled by rare earth-transition metal ferrimagnets holds the promise of ultrafast, energy efficient spintronic technologies. This work investigates laser-induced picosecond spin currents generated by ferrimagnetic GdCo via terahertz emission spectroscopy. A suppression of the THz emission and spin current is observed at magnetization compensation when varying the temperature or alloy composition in the presence of a magnetic field. It is demonstrated that this is due to the formation of domains in the GdCo equilibrium magnetic configuration. Without an applied magnetic field, the picosecond spin current persists at the compensation point. The experimental findings support the model for THz spin current generation based on transport of hot spin-polarized electrons, which is dominated by the Co sublattice at room temperature. Only at low temperature a comparable contribution from Gd is detected but with slower dynamics. Finally, spectral analysis reveals a blueshift of the THz emission related to the formation of magnetic domains close to magnetization compensation.

**1. Introduction**

The field of THz spintronics seeks to harness the spin angular momentum of electrons in the characteristic timescale of their exchange interaction to develop faster, more energy-efficient information technologies.[1–3] One of the principal material platforms in this area are rare earth-transition metal (RE-TM) ferrimagnets.[4–6] In these systems, the net angular momentum and magnetization are determined by the superposition of the antiferromagnetically coupled RE and TM sublattices and, therefore, depend strongly on temperature and composition. At the corresponding compensation temperatures ($T_A$ and $T_M$), where the RE and TM contributions are equal in magnitude, the angular momentum and magnetization vanish.[7] These properties enable RE-TM alloys, or synthetic structures, to display high-frequency dynamics akin to antiferromagnets, while preserving a finite net magnetization that can be easily probed or manipulated, rendering them attractive for THz spintronics applications.[6]

More concretely, a renewed interest on RE-TM magnets surged after the discovery of all-optical helicity-independent switching (AOHIS),[8,9] which has the potential to revolutionize magnetic storage or logic technologies. This effect consists in the reversal of the net magnetization on the picosecond timescale with minimal energy dissipation upon excitation with a femtosecond laser pulse.[2] The first stage in an AOHIS event is the asymmetric ultrafast demagnetization of the RE and TM sublattices,[10] whose underlying mechanisms are an active discussion topic.

In addition to the triggering of ultrafast demagnetization dynamics, a femtosecond laser pulse can produce a THz spin current in multilayer structures.[11–14] Both processes are intrinsically linked, and it was recently demonstrated that in certain systems the exited spin current is proportional to the negative of the time derivative of the magnetization.[15,16] Therefore, by measuring the generated spin current, it is possible to gain a better understanding of the physics underpinning AOHIS.

The origin of these laser-induced spin currents is in itself an open research question. One of the leading theories considers that the pump excites high-energy hot electrons with spin-dependent mobilities and lifetimes that propagate as a superdiffusive spin current away from the magnetic material into an adjacent layer.[12,17,18] Alternatively, the bulk spin pumping model argues that the energy deposited by the laser in the itinerant electrons of the magnet enhances electron-magnon scattering, resulting in a local spin accumulation that can diffuse into a neighbouring material.[13,19,20]

Optically induced spin currents produced by RE-TM ferrimagnets have been investigated through time-resolved magneto-optic Kerr effect (MOKE) spectroscopy. Initial measurements found that certain features in spin accumulation transients could not be accounted for by only considering a spin current produced by the TM sublattice in ferrimagnetic GdFeCo, suggesting a significant Gd contribution.[14] Subsequent experiments on spin valve structures, where the spin current produced by GdFeCo was exploited to switch a nearby ferromagnetic layer in the picosecond timescale, agreed with that conclusion.[21–24] There, the switching was explained in terms of spin angular momentum transfer into the ferromagnetic layer where the spin current produced by Gd ostensibly played a crucial role.[25] However, recent spin transfer torque measurements showed that the RE in a Gd/Co synthetic ferrimagnet can generate a spin current that excites spin waves in a nearby perpendicularly magnetized ferromagnet only at GHz frequencies, not in the THz range.[26]

A similar discrepancy has arisen from THz emission spectroscopy studies of ultrafast dynamics in RE-TM ferrimagnets, where the THz generation was observed to strongly decrease at certain temperatures.[27–29] This has been interpreted as evidence that the RE sublattice can generate a spin current capable of compensating the spin current produced by the TM, cancelling out the net spin generation.[27,29] In contrast, other works have assumed that REs such as Gd and Tb contribute negligibly to the spin current and THz emission.[28,30,31] In the light of the above-mentioned experiments, the characterization of the picosecond spin current produced by a RE-TM ferrimagnet and the detection of a possible RE contribution have become relevant research goals.

In this work, we study THz emission from heterostructures containing GdCo in order to probe laser-induced spin angular momentum transfer. Our results align with the model in which there is an element-specific origin of the picosecond spin current at room temperature. The dependence of the THz signal spectrum on alloy composition indicates that this spin current is dominated by Co. However, at 6 K, we find that the Gd sublattice produces a significant spin current, albeit with a spectrum shifted to lower frequencies compared to that of Co.

Moreover, by investigating the evolution of the THz signal around the magnetization compensation temperature under different magnetic fields, we explain the suppression of the THz emission, and thus of the spin current, with the formation of magnetic domains. We determine that this multi-domain state also affects the THz emission spectrum, in the form of a shift towards higher frequencies.

Our experiments evidence the tunability of RE-TM ferrimagnets as sources of picosecond spin current and highlight the importance of the micromagnetic structure of the alloy on the intensity and spectrum of the THz emission.

## 2. Results and Discussion

Spintronic THz emitters (STEs) were grown by magnetron sputtering in an Ar atmosphere with a pressure of 3 mTorr following the procedure described in reference[32] with the general structure: Intrinsic Si(substrate)/Gd$_x$Co$_{1-x}$($d$)/Cu(2 nm)/HM($\ell$), where the subscripts denote atomic percentages, $d$ is the thickness of GdCo layer, and HM is a heavy metal of thickness $\ell$. In samples 1, 2, and 3, the Gd concentrations are, respectively, $x = 0\%$ (pure Co), 20%, and 100% (pure Gd), with $d = 4$ nm, $\ell = 3$ nm and HM = Pt. Sample 4 has $x = 100\%$, $d = 4$ nm, $\ell = 5$ nm and HM = Ta. For sample 5, a GdCo wedge was grown such that the composition and thickness vary laterally across the sample in the ranges $x \in [5, 50]$ % and $d \in [1.6, 6.4]$ nm, according to **Figure S1**, and HM is Pt with thickness $\ell = 3$ nm. Sample 5 was cut into smaller pieces wherein the Gd concentration changed by about $\Delta x = 5\%$ from edge to edge. In this manner, the interaction between regions with substantially different compositions was eliminated. In the following, we refer to different pieces of sample 5 by the Gd concentration at their centres.

The equilibrium magnetic properties of the samples were characterized by superconducting quantum interference device (SQUID) magnetometry and static MOKE imaging. THz emission spectroscopy was employed to investigate picosecond spin transport. In our experiments, the samples are illuminated with a femtosecond laser pulse that heats up the electron system and drives demagnetization dynamics, concomitantly generating a spin current. The inverse spin Hall effect (ISHE) in the heavy metal layer in the STE converts the injected spin current into a transient charge current that is a source of THz radiation.[12,33] Analysis of the emitted THz pulses therefore provides information about the spin current and magnetization dynamics in the sample in a contactless manner.

In THz emission spectroscopy, the measured quantity, hereafter referred to as the THz signal, is $S(t) = R(t) * E_{\text{THz}}(t)$.[34] Here, $E_{\text{THz}}$ is the projection of the electric field of interest on the axis of the last wire grid polarizer in our setup (see Experimental Section), $*$ denotes convolution, and $R$ is the response function of the spectrometer, which constrained our detection bandwidth to the spectral range 0.3 – 2.8 THz (see **Figure S2**).

The THz electric field is related to the spin current ($j_s$) propagating in the sample via[35]

$$E_{\text{THz}}(\nu) \propto \frac{1}{1 + n + Z_0 \sigma_s(\nu)} j_s(\nu) = Z(\nu)\, j_s(\nu), \qquad (1)$$

where $n = 3.4$ is the virtually frequency-independent THz refractive index of the Si substrate, $Z_0 = 376.7\ \Omega$ the impedance of free space, and $\sigma_s(\nu) = \int_0^{d+\ell} \sigma(y,\nu)\, dy$ the electrical conductivity integrated over the total thickness of the metallic layers (in units of $\Omega^{-1}$). The extraction of the THz conductivity of our samples is reported in Supplemental Material Note S1. To remove the effect of the electrical conductivity from the THz signal, we define the quantity $S^\star(\nu) = S(\nu)/Z(\nu) \propto R(\nu) j_s(\nu)$. Since the response function $R(\nu)$ remains constant throughout our experiments, $S^\star(\nu)$ is useful to monitor relative changes in the spin current spectrum. The deconvolution of the THz signal and extraction of $j_s(\nu)$ is hindered in our setup due to $R(\nu)$, whose bandwidth is too narrow compared to the emission spectrum of a typical STE that can extend over 30 THz[35].

For a STE based on the ISHE, the polarization of the electric field follows from the relationship $\mathbf{E}_{\text{THz}} \parallel \mathbf{j}_c = \theta_{\text{SH}} \mathbf{j}_s \times \hat{\mathbf{S}}$, where $\theta_{\text{SH}}$ is the spin Hall angle of the heavy metal layer, $\mathbf{j}_c$ the charge current, and $\hat{\mathbf{S}}$ the spin polarization direction.[12] The Cu spacers in our samples were added to prevent a large perpendicular magnetic anisotropy,[32] which is detrimental for THz emission because an in-plane magnetization component is required to obtain a THz signal due to the symmetry of the ISHE. Additionally, the Cu layer mitigates a change in the magnetization compensation temperature linked to the magnetic polarization of the HM in proximity to GdCo.[32]

## 2.1. THz emission across the magnetization compensation temperature

In this subsection, we consider the THz emission from sample 2: $Gd_{20}Co_{80}$/Cu/Pt, which has a nominally homogenous Gd concentration. Representative THz pulses measured at 100 K are plotted in **Figure 1**b. The sign change upon reversal of the in-plane magnetic field demonstrates the magnetic character of the signal. The THz electric field is polarized in the direction orthogonal to both the sample plane normal (the propagation direction of the spin current entering the HM) and the bias field, in accordance with the geometry of the ISHE. These symmetries confirm that the measured THz emission follows the phenomenology of a STE described above.[36] For simplicity, hereafter, we only include the emission polarized along the $\hat{\mathbf{x}}$ direction (labelled $S$), according to the reference system in Figure 1a.

Figure 1c shows how the THz emission amplitude changes with temperature for a fixed in-plane magnetic field of 8.5 kOe along $\hat{\mathbf{z}}$. The reported temperatures have been adjusted to account for the effect of DC laser heating (see Supplemental Material Note S2). Two major features can be observed in the waveforms: the polarity of the THz pulses is inverted across 41 K, and the emission vanishes at this temperature, which we denote as $T_{\text{THz}}$. The polarity inversion is a consequence of the switching of the equilibrium sublattice magnetizations induced by the crossing of the magnetization compensation temperature in the presence of a bias magnetic field. From SQUID magnetometry, we determined $T_M = 41.4 \pm 0.3$ K (see **Figure 2**c and Supplemental Material Note S3). Above $T_M$, where the Co magnetization is larger than the Gd magnetization, the former aligns with the external field; below $T_M$, the sublattices are in the opposite configuration with the Gd magnetization aligned with the external field, as illustrated in the inset of Figure 2c.

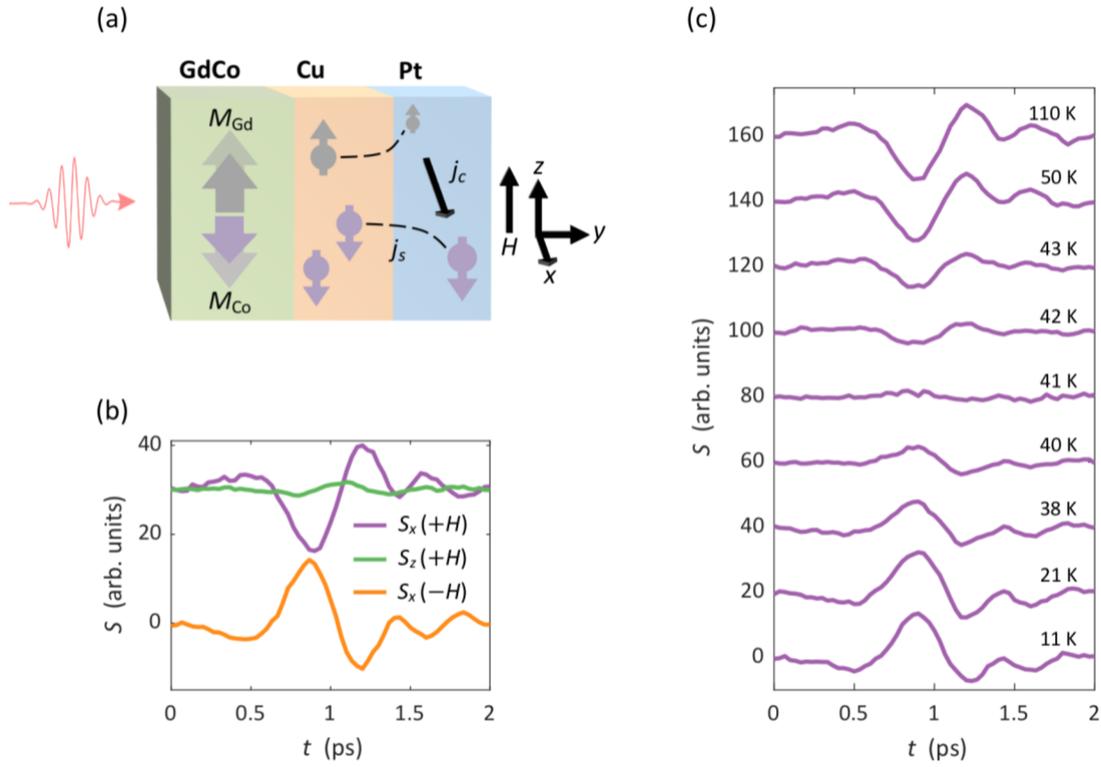

**Figure 1.** THz emission from the homogenous Gd$_{20}$Co$_{80}$(4 nm)/Cu(2 nm)/Pt(3 nm) sample. a) Schematic of the THz emission mechanism. Upon irradiation with a femtosecond laser pump, the magnetic film demagnetizes, and a picosecond spin current pulse is generated, which is then converted into a charge current through the ISHE in the heavy metal layer. This charge transient produces THz radiation that we measure via electro-optic sampling. An in-plane bias magnetic field is applied along the $\hat{\mathbf{z}}$ direction, resulting in THz emission polarized along $\hat{\mathbf{x}}$. b) THz pulses demonstrating the characteristic symmetries of the THz signal in a STE. The measurements were taken at 100 K with $H = 8.5$ kOe. c) Temperature dependence of the THz emission from the Gd$_{20}$Co$_{80}$/Cu/Pt sample measured under a bias field of 8.5 kOe.

In order to understand the vanishing of the THz signal, we investigated how the external magnetic field affects the THz emission intensity, which we quantify with the variable

$$I = \text{sgn}[S(t_{\max})] \int_{t_i}^{t_f} |S(t)| dt, \qquad (2)$$

where sgn[·] is the sign function, $t_{\max}$ the time at which the absolute value of the THz signal has its maximum, and $t_i$, $t_f$ were chosen to encompass completely the THz pulses. In Figure 2a and 2c, we plot the temperature dependence of $|I|$ and compare it with the $\hat{\mathbf{z}}$-component of the static magnetization ($M_z$) measured via SQUID magnetometry under the same bias field of 8.5 kOe. The $M_z$ data is described in detail in Supplemental Material Note S3, and here we simply note that the observed broad valley is caused by the magnetization compensation. As remarked above, the signal intensity plummets at $T_{\text{THz}}$, and this temperature coincides with $T_M$ within the experimental uncertainty.

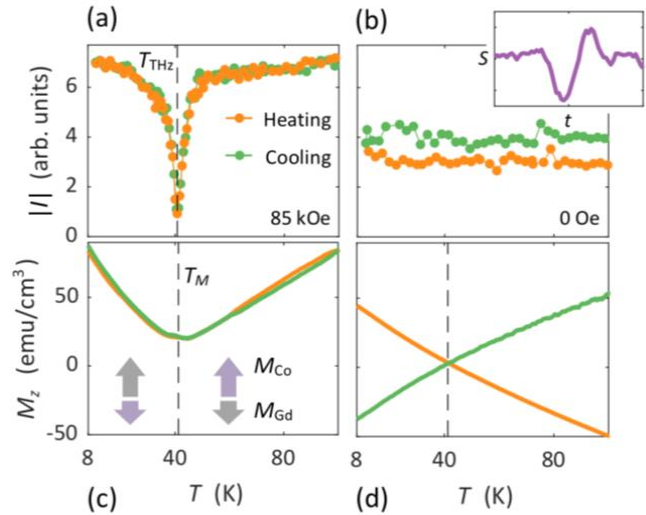

**Figure 2.** THz emission intensity compared to the static magnetization of the Gd$_{20}$Co$_{80}$(4 nm)/Cu(2 nm)/Pt(3 nm) sample as a function of temperature around the compensation point. a) and b) Integrated intensity of the THz signal (defined in Equation 2) at 8.5 kOe and 0 Oe, respectively. Data in orange (green) was taken while heating (cooling). The lack of significant differences in the two thermal cycles shows the stability of the cryostat temperature during the THz measurements. The dashed line in a) corresponds to $T_{\text{THz}}$, the temperature at which the THz intensity is the lowest. The inset in b) is a representative THz pulse detected with zero external magnetic field at 80 K. c) and d) Static magnetization along the $\hat{\mathbf{z}}$ direction measured by SQUID magnetometry under applied fields of 8.5 kOe and 0 Oe, respectively. The dashed lines correspond to the magnetization compensation temperature $T_M = 41.4$, which was obtained by exploiting the thermal hysteresis of the equilibrium magnetization as explained in Supplemental Material Note S3. The inset in c) illustrates that above (below) $T_M$ the Co (Gd) sublattice magnetization is larger than the Gd (Co) one.

On the other hand, we observe a strikingly different behaviour when we apply a magnetic field of 8.5 kOe far away from $T_M$ (6 K when heating and 300 K when cooling), then remove the field and vary the temperature (see Figure 2b and 2d). In this case, the THz emission does not change sign and its intensity remains constant in the vicinity of $T_M$. A representative pulse measured with this field protocol is displayed in the inset of Figure 2b to emphasize that there is a clear THz signal visibly distinguishable from noise. Therefore, considering Equation 1, Figure 2b implies that a finite spin current pulse is excited even when the net magnetization is zero, with a strength that is about half of that in the uncompensated, saturated collinear state.

Prior works on other RE-TM ferrimagnets (GdFeCo and TbFeCo) suggested that the suppression of the THz emission in those systems at a certain temperature is caused by a cancellation of the spin current produced by the antiparallel magnetic sublattices.[27,29] If this was the case for GdCo, the vanishing would occur even when the temperature was varied under zero bias field, which is not seen in Figure 2b. Similar zero field THz emission experiments should be carried out with GdFeCo and TbFeCo STEs to test the hypothesis based on a spin current cancellation.

In contrast, we argue that the suppression of the THz signal in GdCo stems from the loss of the magnetization component along the bias field direction due to the formation of domains and an in-plane canting of the spin polarization. The emergence of a multi-domain structure in sample 2 is supported by the *I* vs *H* hysteresis loops measured at different temperatures and reported in Supplemental Material Note S4. This multi-domain state can be attributed to small variations in the GdCo composition (see Supplemental Material Note S4), which are known to occur in RE-TM alloys because of elemental segregation and lead to spatially dependent magnetization compensation temperatures.[37,38] In particular, the Gd concentration is approximately proportional to the local magnetization compensation temperature,[39] as displayed in **Figure 3**a.

In a cooling measurement, when the temperature is just above the spatially averaged $T_M$, the sublattice magnetizations of the regions with higher Gd content will switch if the applied magnetic field is above the local coercivity. The regions with lower Gd content will not have switched yet, resulting in the coexistence of domains with oppositely aligned sublattice magnetizations. These domains produce THz radiation with different sign, and when the temperature is risen to $T_{\text{THz}}$ their THz emission cancels out. An analogous cancellation occurs when heating, but in that case the Co-rich regions of the sample switch first. However, if the temperature is varied without an applied field, there is no Zeeman energy to drive the switching of the sublattice magnetizations, and, therefore, the micromagnetic structure and the THz emission do not change across $T_M$. At the same time, the THz signal intensity in the absence of a bias field is a fraction of the intensity measured with 8.5 kOe away from $T_M$ because the remanent sublattice magnetizations are smaller than the saturation sublattice magnetizations.

Additional mechanisms may contribute to the vanishing of the THz signal. For instance, we detected an in-plane rotation of ~40° in the THz electric field polarization close to $T_{THz}$ under a low bias field along $\hat{z}$, indicating that the spin polarization tilts away from the field direction (see Supplemental Material Note S5). Due to the geometry of the ISHE, such canting diminishes the recorded signal, which in our case is only the $\hat{x}$-component of the THz electric field. In a macrospin approximation, this tilting can be understood as an intermediate step in the reversal of the Gd and Co macrospins across $T_M$.[40] In reality, the sperimagnetic structure in GdCo adds complexity to this process since the Gd and potentially the Co spin orientations are distributed within cones of varying aperture even away from compensation.[41] Also, a spin-flop transition,[37,40,41] visible in Figure 2c and discussed in Supplemental Material Note S3, can induce a non-collinear magnetic state in the presence of a high magnetic field and further supress the THz emission. In the absence of an external magnetic field, these effects are not active, and the THz emission is unaltered around $T_M$, as in Figure 2b.

Finally, the proximity of $T_M$ and $T_{THz}$ in Figure 2 supports the assertion that the vanishing of the THz emission is caused by the static magnetization state of GdCo and not by dynamical properties as claimed before.[42] For example, in thermodynamic equilibrium, the magnetization dynamics of a ferrimagnet are governed by the angular momentum compensation temperature.[43] Due to the higher Landé factor of the TM compared to the RE,[44] $T_A$ can be tens of kelvin higher than $T_M$,[45] but no distinctive feature away from $T_M$ was seen in our THz emission experiments (see Figure 3b). In the past, significant discrepancies between $T_M$ and $T_{THz}$ have been found for other RE-TM ferrimagnets.[27,29] However, such measurements could have been affected by different temperature calibrations in different equipment, laser heating, and aging of the samples between experiments, which can result in a drift of $T_M$ of tens of Kelvin in a few weeks after deposition.[46]

To further elucidate how local changes in alloy composition lead to the suppression of the THz signal, we investigated the emission from the wedge sample 5 where the Gd concentration is intentionally varied, as described in the next subsection.

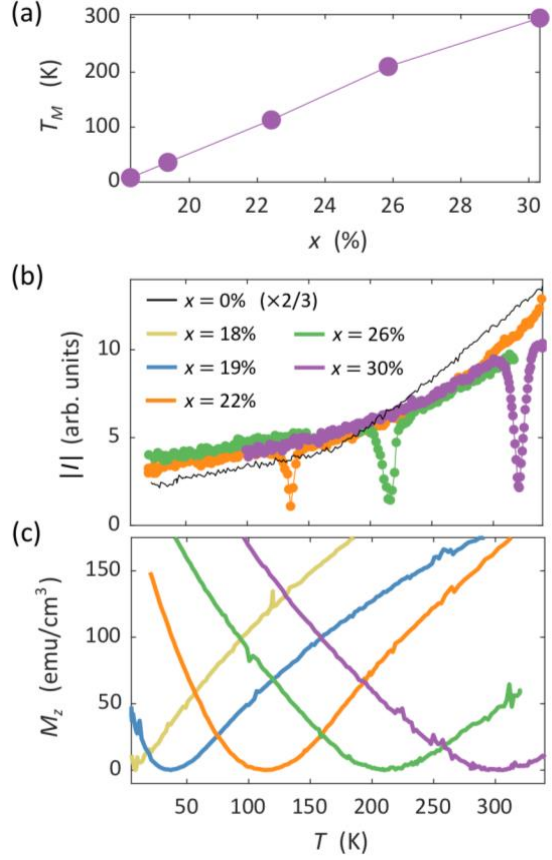

**Figure 3.** THz emission intensity compared to the static magnetization for different GdCo concentrations in the $Gd_xCo_{1-x}(d)/Cu(2\ nm)/Pt(3\ nm)$ sample. a) Magnetization compensation temperature as a function of the $Gd_xCo_{1-x}$ concentration extracted from the SQUID magnetometry data in panel c). The $T_M$ values were obtained by fitting fourth order polynomials to the $M_z$ vs. $T$ curves and taking the position of the minimum of the fitted functions. b) Integrated intensity of the THz signal as a function of temperature for different $Gd_xCo_{1-x}$ concentrations measured under a bias field of 8.5 kOe. The curve corresponding to $x = 0\%$ was reduced by a factor of 2/3 for comparison purposes. c) Temperature dependence of the static magnetization along the $\hat{z}$ direction for different Gd concentrations.

## 2.2. THz emission as a function of Gd concentration

Here, we study the temperature dependence of the THz emission in the composition range $22\% \leq x \leq 30\%$, corresponding to the magnetization compensation temperatures shown in Figure 3a. Figure 3b presents the evolution of the integrated THz intensity measured under an in-plane bias field of 8.5 kOe, similarly to our measurements on the homogenous $Gd_{20}Co_{80}$ structure. For reference, the emission from the Co/Cu/Pt sample is also included. In all cases, the intensity follows an overall decreasing trajectory when lowering the temperature, that we ascribe to the increase of the THz conductivity (see Equation 1 and **Figure S3**) and changes in the spin Hall angle of Pt.[47,48] For a finite Gd concentration, this trend is interrupted by narrow dips that are close to the respective magnetization compensation temperatures, corresponding to the minima in the $M_z$ vs $T$ curves (see Figures 3b and 3c). The discrepancies between $T_{THz}$ and $T_M$ are attributed to the different sample areas probed in the two techniques: spots with a diameter of 400 μm in THz spectroscopy corresponding to fairly constant compositions; and ∼5 × 5 mm$^2$ regions in SQUID magnetometry which yielded averaged properties over a range of $\Delta x \approx 5\%$.

Next, we compare these temperature-induced changes with the dependence of the THz emission on Gd concentration at fixed temperature. In order to remove the effect of the different THz conductivities (see Figure S3), we focus on the normalized signal $S^\star$, defined at the beginning of Section 2. As part of our analysis, we calculate the integrated intensity of $S^\star$, labelled $I^\star$, as in Equation 2 with $S$ replaced by $S^\star$.

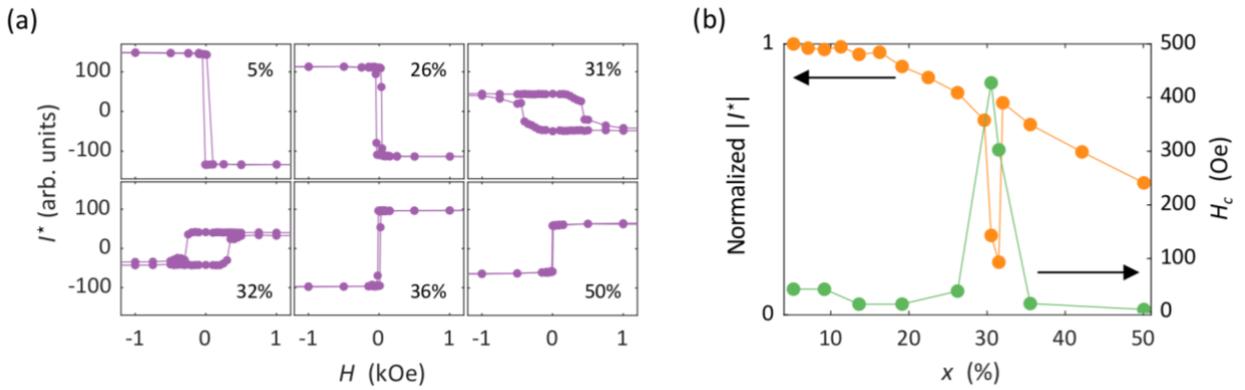

**Figure 4.** Integrated THz intensity $I^\star$ for different GdCo concentrations of the Gd$_x$Co$_{1-x}$(d)/Cu(2 nm)/Pt(3 nm) sample at room temperature. a) Integrated intensity ($I^\star$) hysteresis loops across the compensation composition. b) Integrated THz intensity $I^\star$ at 8.5 kOe (orange) and coercivity of the integrated intensity hysteresis loops (green) as a function of Gd atomic percentage. The intensity $I^\star$ was calculated by integrating the absolute value of the $S^\star(t)$ as in Equation 2. The shown intensities were normalized with respect to value at $x = 5\%$.

**Figure 4**a presents $I^\star$ vs $H$ hysteresis loops for different Gd concentrations measured at room temperature. The change in the polarity of the loops from $x = 31\%$ to 32% reflects the transition from a Co-rich to a Gd-rich alloy magnetization. The decrease in their height corresponds to the vanishing of the THz signal, which is also manifested as the dip in Figure 4b. There, we juxtapose the intensity $I^\star$ obtained at 8.5 kOe with the coercive fields $H_c$ extracted from the hysteresis loops in Figure 4a. The intensity minimum is in correspondence

with the coercivity maximum, and the latter occurs along the compensation line in the $(x,T)$ phase diagram.[38] Hence, through Figure. 3 and 4, we have shown that the photo-excited spin current from GdCo is supressed at magnetization compensation, regardless of whether this is reached through temperature or composition variation.

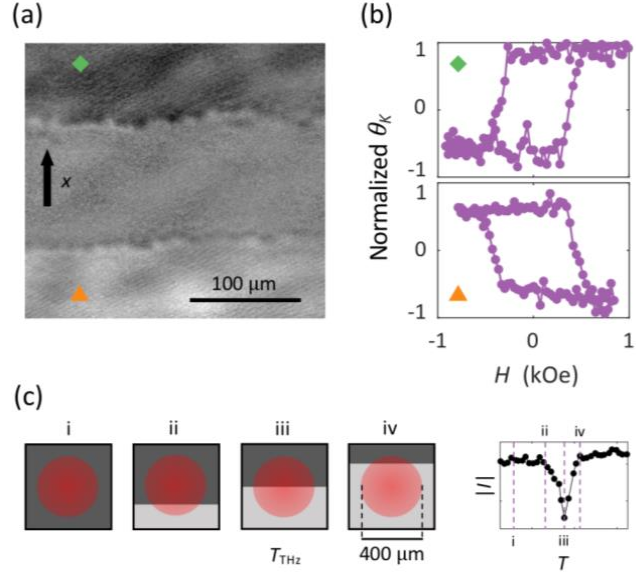

**Figure 5.** Static MOKE characterization of the sample $Gd_{31}Co_{69}$(2.9 nm)/Cu(2 nm)/Pt(3 nm). a) MOKE imaging of the region in the sample with magnetization compensation temperatures in the vicinity of room temperature. The composition gradient is such that the Gd concentration increases in the vertical direction, as indicated by the black arrow. The light (dark) contrast represents domains where the Co (Gd) sublattice magnetization is aligned with the bias field of $-1.8$ kOe. The grey intermediate region corresponds to an unsaturated area where the coercive field was larger than the bias field. b) Kerr rotation ($\theta_K$) measured at 250 μm above and below the centre of the image in a), as symbolized by the green and orange markers. The Kerr hysteresis loops were obtained by averaging the MOKE contrast in regions of size ~450 μm by 150 μm. c) Sketch of the changes in the static magnetization configuration of the area probed in the THz experiments at the temperatures indicated in the THz intensity plot on the right. The light and dark regions represent domains where the Co and Gd sublattices are dominant, respectively. The red circle depicts the pump laser spot.

For $x \approx 31\%$ with magnetization compensation at room temperature, we imaged the multi-domain structure responsible for the suppression of the THz signal. A representative static MOKE micrograph taken with $H = -1.8$ kOe is shown in **Figure 5**a, where the composition gradient of the sample is shown by the back arrow along the vertical direction. The MOKE signal is mostly sensitive to the Co sublattice,[49,50] so the well-defined light and dark contrast regions are Co-rich and Gd-rich domains, respectively, where the Co magnetization points in opposite directions. This is confirmed by the hysteresis loops in Figure 5b, which were taken 250 μm above and below the centre of the MOKE image. The grey middle area corresponds to an unsaturated region where the bias field was below the coercive field. This unsaturated stripe is expected to have a width of a few tens of microns in the THz experiments when the applied field is 8.5 kOe (see **Figure S8**), and thus it plays a minor role in the THz emission data.

Considering that the pump spot size has a FWHM of 400 μm in the THz measurements, it becomes clear that the THz signal vanishes when the pump illuminates equal amounts of Gd-dominated and Co-dominated domains, where the local net magnetizations are opposite. Furthermore, the border between the two domains shifts with temperature due to the spatial variations of $T_M$, so that sufficiently away from compensation a single domain is probed, as sketched in Figure 5c.

**2.3 Gd contribution to the THz signal**

In the previous subsections, we established that the vanishing of the THz emission is not a consequence of the cancellation of the spin current produced by the RE and TM sublattices. Here, we explain why these two contributions to the spin current cannot fully offset each other by examining their relative strengths and timescales through the analysis of the THz emission spectrum as a function of the alloy composition.

**Figure 6**a and 6c display the normalized THz signal and its corresponding Fourier transform for different Gd concentrations measured at room temperature with a bias field of 8.5 kOe. The similarity amongst the spectra for the extreme studied compositions $x = 0\%$ (sample 1), 5% and 50% (sample 5) suggest that the picosecond spin currents in these samples share the same origin. The different behaviour for $x = 32\%$ is explained below, in the discussion of **Figure 7**. Additionally, note that, in the sample with $x = 50\%$, the Co magnetization sublattice is antiparallel to the magnetic field because its Gd concentration is well above the room temperature compensation value $x_M \approx 31\%$. Thus, the opposite sign and nearly identical spectrum of the THz emission for $x = 0\%$ and 50% indicate that, at room temperature, the spin current is dominated by the Co sublattice in a broad composition range.

Figures 6b and 6d present analogous data measured at 6 K. At this temperature, elemental Gd is in its ferromagnetic phase, and therefore it is possible to detect a THz signal from samples 3 and 4. In the past, THz emission from Gd-based heterostructures has been ascribed to the anomalous Hall effect in this RE.[31] We ruled out that this is the main emission mechanism in our STEs by comparing the THz pulses generated by Gd/Cu/Pt and Gd/Cu/Ta (see **Figure S9**a). The inverted polarity of the obtained waveforms, stemming from the different signs of the Pt and Ta spin Hall angles, verifies that the signal originates from the combination of spin current generation and the ISHE, as described at the beginning of Section 2. Further confirmation of the relationship between the Gd equilibrium magnetization and its THz emission is provided in Figure S9.

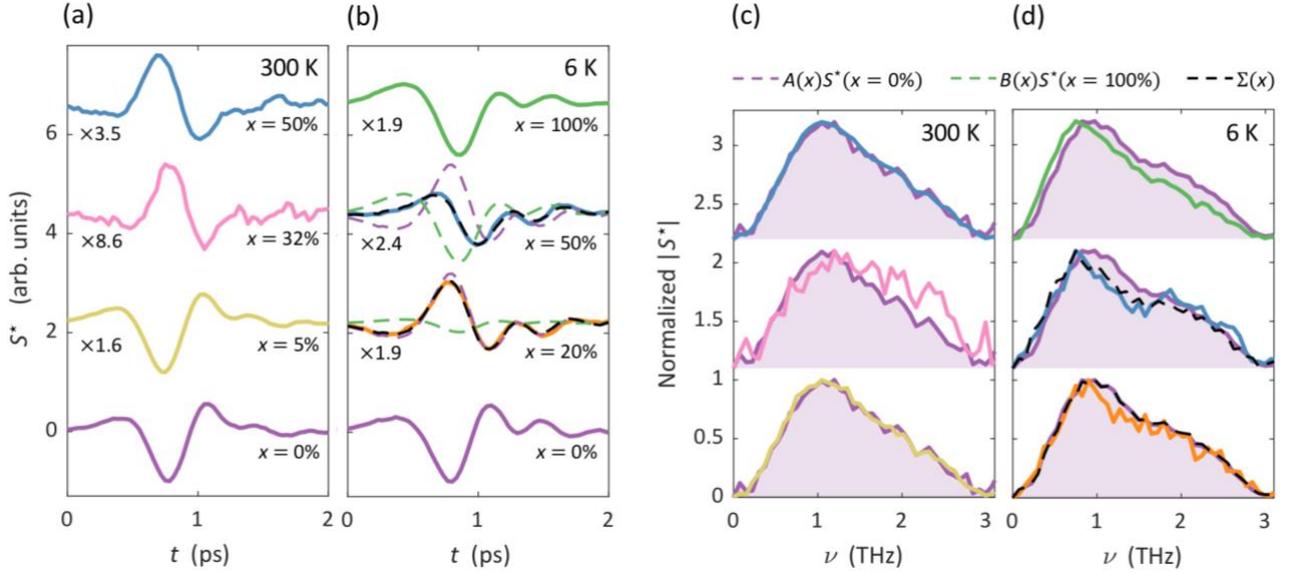

**Figure 6.** Contribution of the Gd sublattice to the THz signal. a) and b) Normalized THz signal measured for different Gd concentrations under a bias field of 8.5 kOe at 300 K and 6 K. The data for $x = 5\%$, 20%, 32%, and 50% were obtained using the $Gd_xCo_{1-x}(d)$/Cu(2 nm)/Pt(3 nm) sample. The data for $x = 0\%$ and 100% were collected using the Co(4 nm)/Cu(2 nm)/Pt(3 nm) and Gd(4 nm)/Cu(2 nm)/Pt(3 nm) samples, respectively. The multiplicative factors on the left of the panels indicate by how much the waveforms were rescaled to have the same amplitude as the $x = 0\%$ pulse at the corresponding temperature. c) and d) Absolute value of the Fourier transforms of the transients $S^\star(t)$ in a) and b). The Co/Cu/Pt spectrum (in purple) is repeated for comparison purposes. The spectra are normalized so that their peak amplitude is equal to one. In all panels, the solid lines are experimental data. In b) and d), the black dashed lines correspond to linear combinations of the Gd/Cu/Pt and Co/Cu/Pt signals. In b), the green (purple) dashed lines represent the Gd/Cu/Pt (Co/Cu/Pt) component of each superposition. The amplitudes of these components are $A(x = 50\%) = -0.41$, $B(x = 50\%) = 0.74$, $A(x = 20\%) = -0.52$, and $B(x = 20\%) = 0.17$.

The spectra in Figure 6d indicate that the spin current produced by pure Gd has an overall slower frequency content than that produced by Co. Considering the connection between THz spin transport and ultrafast demagnetization dynamics,[15,16] the shift between the Co and Gd spectra is consistent with the different demagnetization time scales in the two elements, namely, ~200 fs for Co, ~1 ps for the first demagnetization stage of Gd, and tens of picoseconds for its second stage.[51]

The THz emission spectrum of the alloy at low Gd concentrations, such as $x = 20\%$, closely resembles the THz emission from Co. As the Gd concentration is increased, a suppression of the Fourier components between about 0.9 and 1.7 THz develops, as illustrated in the data for $x = 50\%$, evidencing the existence of a nonnegligible Gd contribution to the picosecond spin current. Indeed, we are able to approximately reproduce the THz emission from GdCo with a linear combination of the signals from pure Gd and Co (dashed lines in Figure 6) that can be expressed as $\Sigma(t,x) = A(x)S^\star(t, x = 0\%) + B(x)S^\star(t, x = 100\%)$. In this superposition, the ratio of the amplitude of the Gd component ($B$) to that of the Co component ($A$) increases with the Gd concentration. Hence, for high Gd content and at a sufficiently low temperature, the Co sublattice no longer dominates the optically induced spin current generation, and Gd can produce a comparable contribution.

The temperature dependence of the Gd contribution can be understood from the Gd electronic band structure. Most of the magnetic moment of this element (7 $\mu_B$ per atom)[52] lies in localized $4f$ orbitals that cannot be directly excited with the comparatively low energy of the pump (1.5 eV for our laser, versus the potential energy of approximately 8.4 eV of the $4f$ spins).[53] However, the pump pulses can be absorbed by $5d$ valence electrons (carrying 0.55 $\mu_B$ per atom),[52] which creates a hot electron spin current and increases the frequency of scattering events between the localized and itinerant spins.[11] The latter effect, termed bulk spin pumping,[19] creates a spin accumulation in the $5d$ subsystem that can diffuse into the subsequent layers of the STE, constituting an additional spin current generation mechanism.

Assuming a similar behaviour in the alloy, the room temperature contribution of the Gd sublattice to the detected spin current is negligible for two reasons. Firstly, the characteristic timescale at which $4f$ localized spins and itinerant spins scatter is too slow to be efficiently probed with our detection bandwidth of 0.3 – 2.8 THz; for example, this scattering time has been claimed to be 20 ps for Gd in a Gd/Co bilayer.[26] In contrast, the response function of our setup attenuates the sensed THz emission almost linearly with respect to frequency between 0 and 1 THz, so that Fourier components below 0.3 THz are strongly supressed (see Supplemental Material Figure S2). At frequencies below our detection threshold, the room temperature Gd contribution might be more sizable, as indeed observed in previous work.[26] Secondly, the spin current transported by hot electrons is weak due to the small spin polarization of the 5$d$ bands. Nevertheless, at low temperature, this spin polarization increases significantly[54] and the ultrafast demagnetization dynamics of Gd speed up,[55] potentially signalling an acceleration of its laser-induced spin current. Consequently, the Gd contribution to the THz spin current becomes more prominent with decreasing temperature.

Note that the analysis represented in Figure 6 that considers the GdCo THz emission as a linear combination of the Co and Gd signals assumes that the spin dynamics are the same in the alloy and in the individual elements, which, in general, is not true.[51,56] Nonetheless, Figure 6 correctly highlights that the spin current produced by the RE and TM sublattices are associated with different characteristic time scales. Thus, there is no "THz compensation temperature" at which the Gd and Co contribution can completely cancel out, in agreement with the previous subsections.

## 2.4. THz emission spectrum close to compensation

As described above, the Gd sublattice has a significant indirect impact on the THz emission through its influence on the GdCo static magnetic configuration. In particular, we have discussed how the formation of magnetic domains supresses the THz signal. In the following, we describe how this domain structure can modify the THz emission spectrum close to magnetization compensation. Figure 7a, corresponding to sample 5 with $x \approx 25\%$, shows that sufficiently far away from $T_{\text{THz}} = 181$ K the normalized THz transients resemble single-cycle pulses, but close to $T_{\text{THz}}$ the emission exhibits additional oscillations. In the frequency domain, this is translated into a marked shift of the peak in the spectra to higher frequencies close to compensation (Figure 7b).

A more detailed Fourier dataset is provided in Figure 7c and 7d, displaying nearly constant spectra when heating up towards the magnetization compensation temperature. After crossing $T_M$, which is represented by the $\pi$ phase shift in Figure 7d, there is a sudden blueshift in the amplitude spectra that fades away with increasing temperature. We also observe a similar spectral change at room temperature when the Gd concentration is close to its compensation value, as evidenced by the data for $x = 32\%$ in Figure 6b.

As explained in the previous subsections, domains with opposite alignment of the magnetic sublattices coexist around the magnetization compensation temperature. We ascribe the change in the THz emission spectrum close to $T_M$ to the interference of the spin current produced by these magnetic domains. This is supported by the fact that we could reproduce the THz emission just above $T_M$ with a superposition of the signals above and below compensation (dashed lines in Figure 7a and 7b), where the two summands represent the two types of magnetic domains.

The modified spectrum of this superposition arises from a time shift of about 30 fs between the THz emission of sample 5 above and below $T_M$. This delay across compensation is revealed in Figure 7d, where we plot the time corresponding to the maximum of $|S^\star(t)|$ (labelled $t_{\max}$) as a function of temperature. Our estimation of the magnitude of the time shift is hindered by the duration of our probing laser pulse (50 fs), but its existence is confirmed by its consistent detection sufficiently far from $T_M$. The repeatability of this effect was verified in various pieces taken from sample 5 with different compositions and magnetization compensation temperatures.

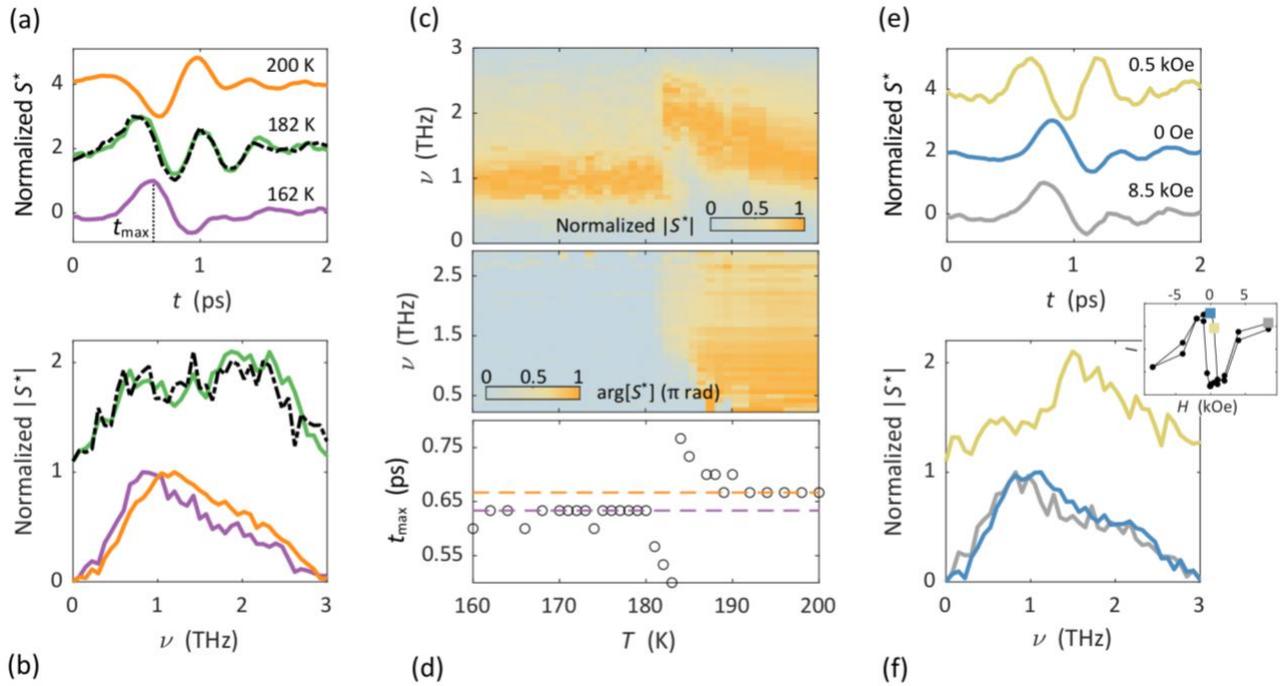

**Figure 7.** THz emission spectra across magnetization compensation. a) Normalized THz pulses emitted by sample $Gd_{25}Co_{75}$/Cu/Pt at temperatures away (200 K, 162 K) and close (182 K) to the magnetization compensation point, measured with a bias field of 8.5 kOe. The vertical dotted line indicates the time ($t_{\max}$) corresponding to the maxima of $|S^\star(t)|$ at 162 K. b) Absolute value of the Fourier transforms of the transients in a). The black dashed lines in a) and b) correspond to a linear combination of the emission at 200 K and 162 K. c) Norm and phase of spectra from sample $Gd_{25}Co_{75}$/Cu/Pt taken at temperatures ranging from 160 K to 200 K. d) Time ($t_{\max}$) corresponding to the maxima of the absolute value of the $S^\star(t)$ data used to obtain c). The dashed purple and orange lines underscore the shift of these times across $T_M$, which corresponds to one experimental time step (33 fs). e) Normalized THz signal from the homogenous sample $Gd_{20}Co_{80}$/Cu/Pt obtained at 38 K with bias fields 8.5 kOe, 0 kOe, and 0.5

kOe. The two latter measurements were taken in the branch of the hysteresis loop where the magnetic field was increased, as indicated in the inset by the coloured square markers. f) Absolute value of the Fourier transforms of the pulses in e).

We leave the determination of the origin of this small time delay for future work. We hypothesize that it could be related to the sperimagnetism of GdCo,[41] which tilts the sublattice spins asymmetrically across $T_M$ in a manner reminiscent of both Figure 7c and the spin reorientation reported in Figure S7. Considering that in certain systems the amplitude of the equilibrium magnetization is coupled to the characteristic time scale of the laser-induced spin current,[16] an uneven canting of the GdCo magnetic sublattices may alter the spin transport dynamics across $T_M$. An alternative explanation could be associated with spin mixing, which has been observed to delay the onset of ultrafast demagnetization in pure Gd by up to 200 fs depending on the equilibrium temperature.[57]

Lastly, a comparable suppression of the lower frequency components of the THz emission was detected in sample 2, which has a nominally homogeneous GdCo concentration, close to $T_M$ when the bias magnetic field is approximately $H_c$ (see Figure 7e and 7f). In this case, the role of the coercive field, determined from the THz hysteresis loop in the inset of Figure 7f, is to promote the formation of a multi-domain structure. As in sample 5, a modified spectrum is obtained due to the interference of the THz emission from different magnetic domains.

## 3. Conclusions

We have conducted a systematic characterization of laser-induced picosecond spin transport in GdCo/Cu/Pt heterostructures through THz emission spectroscopy. In our experiments, the THz signal is supressed when crossing the magnetization compensation line in the composition-temperature phase diagram under a bias magnetic field. We explain this behaviour as a consequence of the GdCo equilibrium magnetization configuration, which partitions into domains close to compensation in the presence of a bias field, due to composition inhomogeneities. Without the external magnetic field, we detected THz emission around and at $T_M$ with an intensity comparable to that in the saturated state, demonstrating that a ferrimagnet can generate a picosecond spin current even when it is fully compensated. The GdCo micromagnetic structure can also affect the shape of the emitted THz pulses. This is associated with a blueshift of the THz emission spectra close to magnetization compensation when the THz signal from different magnetic domains interfere.

At room temperature, we determined that the optically triggered spin current is dominated by the Co sublattice in the explored frequency range, i.e., 0.3 − 2.8 THz. This is consistent with previous studies which claim that there is a significant Gd contribution[14,21–23] since, at room temperature, the Gd spin current occurs at GHz frequencies outside our detection window.[26] On the other hand, at 6 K, we observed that the Gd sublattice can produce a THz spin current commensurate with that generated by Co. Nevertheless, the Gd and Co spin currents cannot fully cancel out at all times because they are associated with different characteristic time scales, which parallels the asymmetric ultrafast demagnetization dynamics of the sublattices.[56]

This work presents a path to reconcile two seemingly incompatible phenomena in the THz dynamics of RE-TM ferrimagnets: the dominant role of the TM in the photo-induced THz spin current at room temperature and the vanishing of the THz emission, previously thought to be a sign of equally strong THz spin generation by the RE and TM sublattices. In this way, our results build on our understanding of the ultrafast magnetization dynamics in RE-TM systems and are relevant for the physical description and development of optically switchable spintronic devices based on spin transfer.

## 4. Experimental Section

*MOKE measurements:* The MOKE experiments were done at room temperature in a longitudinal configuration (incidence angle of 45°) with a pulsed laser repetition rate of 100 kHz, central wavelength of 515 nm, and duration of 150 fs. An in-plane magnetic field ranging from 0 to 2 kOe was applied.

*THz emission spectroscopy:* The THz spectroscopy setup, which is different than the one used for the MOKE experiments, is described in detail in reference[58]. Briefly, it is based on a Ti:Sapphire pulsed laser operating at a repetition rate of 5 kHz, with a central wavelength of 800 nm, and pulse duration of approximately 50 fs. The linearly polarized laser pulses were collimated to a spot with 400 μm of FWHM, corresponding to an incident fluence of 10.3 mJ/cm$^2$. The pump impinged the STEs at normal incidence on the heavy metal side (see experimental geometry in Figure 1a). The samples were placed in a closed-cycle cryostat between the poles of an electromagnet capable of applying in-plane magnetic fields between 0 – 8.5 kOe.

The emitted THz pulses were collected with off-axis parabolic mirrors, filtered with wire grid polarizers that enable polarimetry capabilities, and focused on a 1 mm thick ZnTe detection crystal, where they overlapped with gating laser pulses. The THz electric field was sensed through electro-optic sampling.[34]

**Acknowledgements**

C.C. acknowledges a Royal Society Research Fellowship, a Leverhulme Trust Research Project grant (RPG-2023-271), and a UKRI Frontier Research Guarantee grant (EP/Z000637/1). J.B. acknowledges support from the Royal Society through University Research Fellowships. This project has received funding from the European Union's Horizon 2020 research and innovation programme under the Marie Skłodowska-Curie Grant Agreement No. 861300 (COMRAD). We acknowledge support by the ANR through the UFO project (ANR-20-CE09-0013), the SLAM project (ANR-23-CE30-0047), and through the France 2030 government grants PEPR Electronic EMCOM (ANR-22-PEEL-0009), PEPR SPIN (ANR-22-EXSP-0002 / ANR-22-EXSP-0008), and the MAT-PULSE-Lorraine Université d'Excellence project (ANR-15-IDEX-04-LUE).

Supporting Information

**Origin of the laser-induced picosecond spin current across magnetization compensation in ferrimagnetic GdCo**

*Guillermo Nava Antonio, Quentin Remy, Jun-Xiao Lin, Yann Le Guen, Dominik Hamara, Jude Compton-Stewart, Joseph Barker, Thomas Hauet, Michel Hehn, Stéphane Mangin\*, and Chiara Ciccarelli\**

*Corresponding authors: cc538@cam.ac.uk, stephane.mangin@univ-lorraine.fr

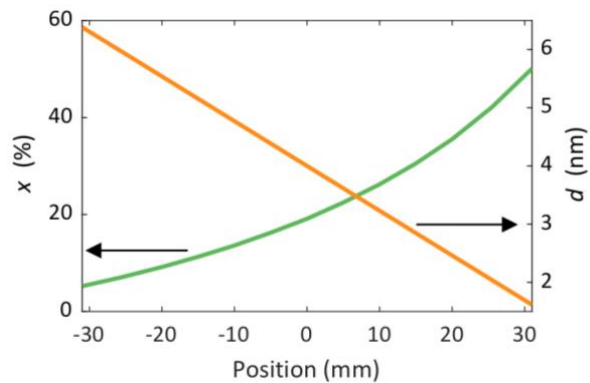

**Figure S1.** GdCo composition (green curve) and thickness (orange curve) as functions of lateral position of sample 5 ($Gd_xCo_{1-x}(d)$/Cu(2 nm)/Pt(3 nm)) calculated from the Co and Gd deposition rates. The variable $x$ denotes Gd atomic percentage. The composition and thickness are nominally constant along the direction on sample 5 perpendicular to that reported on the horizontal axis of this plot.

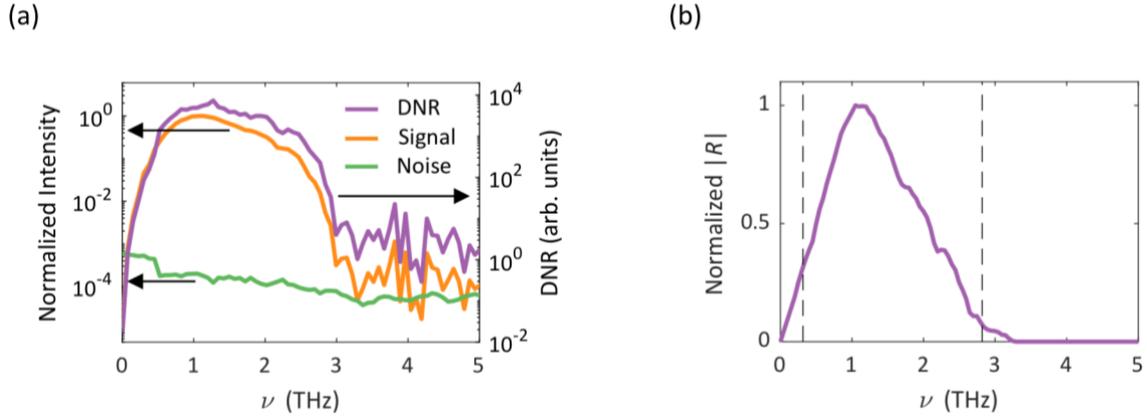

**Figure S2.** Frequency dependence of the sensitivity of the THz emission spectroscopy measurements. a) Dynamic range (DNR) for a typical experiment, quantifying the usable bandwidth of the THz spectrometer. The DNR is defined as the ratio of the Fourier transform of the THz emission from the sample under study to the Fourier transform of the experimental noise.[1] In this plot, the former (labelled as "Signal") was measured using sample 1 (Co(4 nm)/Cu(2 nm)/Pt(3 nm)) at room temperature under a bias field of 8.5 kOe along $\hat{z}$, and the latter was obtained by recording the signal with the laser pump blocked. b) Response function of our THz spectroscopy setup determined experimentally using a ZnTe reference THz emitter, following the procedure outlined in reference[2]. A rectangular filter was applied with cutoff frequency at 3.25 THz, where the DNR shown in a) drops to approximately 1. The response function is dictated by the nonlinear optical properties of the used ZnTe detection crystal, the duration and spatial profile of the laser pulses, and the propagation of the THz radiation from the sample to the detector position. The dashed lines at approximately 0.3 THz and 2.8 THz mark the frequency range where the DNR is greater than 100.

## Supplemental Note S1. THz conductivities of studied samples

We determined the THz stack conductivities ($\sigma_s$, defined in Section 2 of the main text) of our samples via THz transmission experiments. These were carried out in the same setup used for THz emission spectroscopy, following a methodology inspired by[3] and using a 1 mm thick ZnTe crystal as a source of probing THz pulses. To obtain the stack conductivities at room temperature, we measured THz pulses transmitted through the sample of interest ($S_{\text{sample}}(\nu, T = 300 \text{ K})$) and through a Si substrate nominally equivalent to that of the sample ($S_{\text{Si}}(\nu)$). It can be shown through wave propagation analysis, in the thin-film approximation, that

$$\sigma_s(\nu, 300 \text{ K}) \approx \frac{1+n}{Z_0}\left[-1 + \exp\left(\frac{2\pi\nu i \Delta L}{c}(n-1)\right)\frac{S_{\text{Si}}(\nu)}{S_{\text{sample}}(\nu, 300 \text{ K})}\right], \qquad (S1)$$

where $Z_0 = 376.7 \, \Omega$ is the impedance of free space, $c$ the speed of light in vacuum, and $n$ the THz refractive index of silicon. By means of conventional THz transmission spectroscopy,[1] we measured the latter to be $n = 3.4$ and virtually independent of frequency (in our bandwidth) and temperature, in agreement with the literature.[4] In writing Equation S1, we considered that the thickness of the reference Si substrate ($L_{\text{Si}}$) can be different than that of Si substrate of the sample of interest ($L_{\text{sample}}$) by an amount $\Delta L = L_{\text{sample}} - L_{\text{Si}}$. Such a thickness mismatch can lead to artifacts in the phase of the extracted complex conductivity.

In most metals, the room temperature conductivity in the frequency range of interest to us is a real quantity because, in the Drude picture, the electron scattering rate is typically of the order of tens of THz.[3,5] Taking this into account, we calculated $\sigma_s$ in two steps. First, we took $\Delta L = 0$ and substituted the experimentally measured $S_{\text{sample}}$ and $S_{\text{Si}}$ in Equation S1 to obtain $\sigma_{s,0}(\nu, 300 \text{ K}) \in \mathbb{C}$. Then, we enforced $\sigma_s(\nu, 300 \text{ K}) \in \mathbb{R}$ by simultaneously fitting, through the least-squares method, $\text{Re}[\sigma_s(\nu, 300 \text{ K})]$ to $|\sigma_{s,0}(\nu, 300 \text{ K})|$ and $\text{Im}[\sigma_s(\nu, 300 \text{ K})]$ to zero using $\Delta L$ as a fitting parameter. The obtained values for $\Delta L$ were substituted back in Equation S1, together with $S_{\text{sample}}$ and $S_{\text{Si}}$, to yield the final result for the stack conductivity. For all samples, $\Delta L$ never exceeded 5 µm and led to relatively small corrections for $\sigma_{s,0}(\nu, 300 \text{ K})$.

The extracted room temperature stack conductivities were nearly independent of frequency so, for simplicity, we fitted them to constants. Subsequently, we determined the temperature dependence of the stack conductivities by measuring $S_{\text{sample}}(\nu, T)$. Wave propagation analysis, in the thin-film regime, can be applied to find

$$\sigma_s(\nu, T) \approx \frac{1+n}{Z_0}\left(\frac{S_{\text{sample}}(\nu, 300 \text{ K})}{S_{\text{sample}}(\nu, T)} - 1\right) + \sigma_s(\nu, 300 \text{ K})\frac{S_{\text{sample}}(\nu, 300 \text{ K})}{S_{\text{sample}}(\nu, T)}, \qquad (S2)$$

where the previously obtained room temperature stack conductivity should be used. In this instance, an imaginary part of $\sigma_s(\nu, T)$ was allowed since the electron scattering rate can drop substantially with lowering temperature.[5]

The results of applying the procedure described above are condensed in **Figure S3**. There, we use the notation $\sigma_1 = \text{Re}[\sigma_s]$ and $\sigma_2 = \text{Im}[\sigma_s]$. In all samples, an imaginary part of the conductivity develops as the temperature is decreased, which is almost monotonously increasing with frequency. At the same time, lowering the temperature gives rise to the emergence of a negative slope in the real part. Both trends are consistent with the Drude model with electron scattering rates higher than the maximum probed frequency, i.e., $\sim$2.3 THz.

The stack conductivities of the samples containing the $Gd_xCo_{1-x}$ alloy with $x \neq 0\%$ or 100% change continuously from the values in Figure S3d, corresponding to $x = 5\%$, to those in Figure S3e, corresponding to $x = 50\%$. This is partly determined by the position-dependent GdCo thickness of sample 5 shown in Figure S1. The actual conductivity (in units of $1/(\Omega\,\text{m})$) of the GdCo layer may be weakly dependent on Gd concentration for $x \in [5, 50]\%$, similarly to the case of GdFe.[6]

Before the extracted temperature-dependent stack conductivities were used to calculate the effective impedances $Z$ defined through Equation 1 of the main text, we fitted them with third-degree polynomials. In this manner, we were able to work with smooth functions that could be evaluated at any desired frequency within the range in Figure S3.

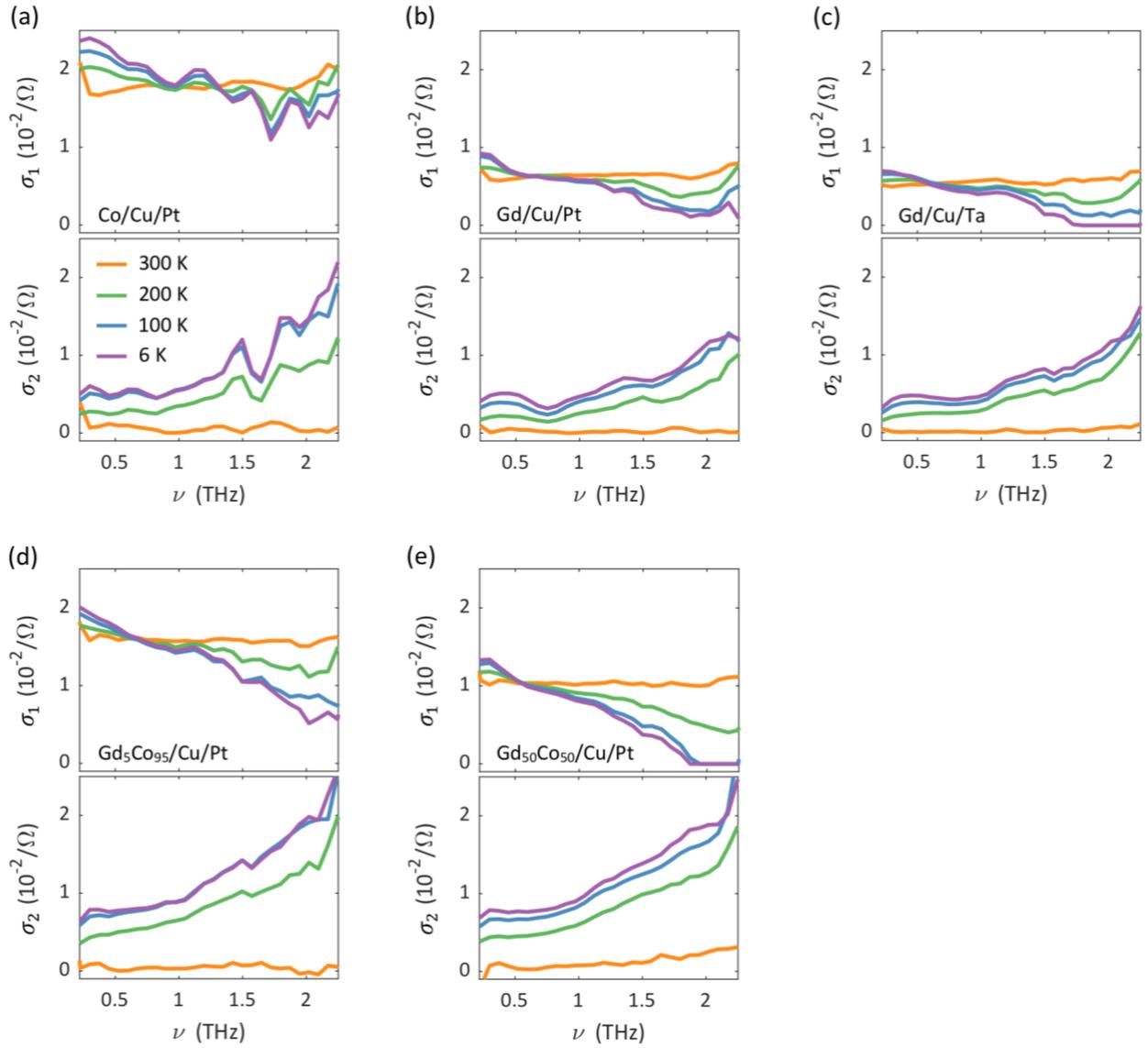

**Figure S3.** Stack conductivities of the studied samples at selected temperatures. a)-c) Conductivities of samples 1, 2, and 4 respectively. d) and e) Conductivities of the extreme compositions in sample 5.

**Supplemental Note S2. Equilibrium temperature correction due to DC laser heating**

We used a relatively high pump fluence of 10.3 mJ/cm² in our THz emission measurements to produce THz pulses considerably stronger than the experimental noise (see Figure S2), which resulted in an increase of the equilibrium temperature of the samples. We took this DC laser heating into account when comparing the THz experiments with the static magnetization data measured by SQUID magnetometry where no such heating was present.

This correction was done by introducing the adjusted sample temperature in the THz experiments $T' = T_0 + \Delta T(F)$, where $T_0$ is the equilibrium temperature when the pump laser was blocked, $\Delta T$ the increment due to laser heating, and $F$ the pump laser fluence. Additionally, we assumed that $\Delta T$ is independent of $T_0$ and that, below a certain threshold fluence $F_{th}$, the DC heating effect is negligible as it induces a temperature shift smaller than the resolution of our temperature sensor.

We determined $\Delta T$ by measuring the integrated THz intensity ($I$) as a function of the cryostat temperature ($T_0$) with different pump fluences and exploiting the sharp minimum of $I$ at $T_{THz}$ (see Figure 2). As highlighted by the rightmost dashed line in **Figure S4**, the minimum of the integrated THz intensity occurs at a fixed temperature when the fluence is low ($F = 3.3$ or $0.8$ mJ/cm²). However, owing to the above-mentioned DC laser heating, the minimum of $I$ shifts to a lower $T_0$ when the fluence is high ($F = 10.3$ mJ/cm²). Thus, by considering that $F_{th} \geq 3.3$ mJ/cm², we concluded that $\Delta T(F = 10.3$ mJ/cm²) is equal to the separation between the dashed lines in Figure S4, namely, 2.7 K. Finally, in the main text, we dropped the notation $T'$ for clarity and simply used the symbol $T$.

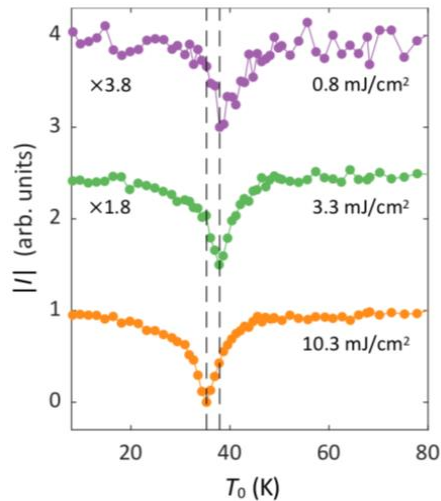

**Figure S4.** Temperature dependence of the integrated THz intensity of the Gd$_{20}$Co$_{80}$(4 nm)/Cu(2 nm)/Pt(3 nm) sample measured with incident pump fluences of 10.3, 3.3, and 0.8 mJ/cm² under an external magnetic field of 8.5 kOe along $\hat{\mathbf{z}}$. The multiplicative factors on the left indicate by how much each data set has to be rescaled so that the corresponding intensity at 80 K has the same value as the intensity at 80 K when $F = 10.3$ mJ/cm². Additionally, the data for different fluences are vertically shifted for clarity. The leftmost and rightmost dashed lines mark $T_{\text{THz}}(F = 10.3 \text{ mJ/cm}^2)$ and $T_{\text{THz}}(F = 3.3 \text{ mJ/cm}^2) \approx T_{\text{THz}}(F = 0.8 \text{ mJ/cm}^2)$, respectively.

**Supplemental Note S3. Determination of magnetization compensation temperature**

The magnetization compensation temperature of sample 2 (Gd$_{20}$Co$_{80}$(4 nm)/Cu(2 nm)/Pt(3 nm)) was determined by measuring its in-plane magnetic moment as a function of temperature under various fixed magnetic fields along $\hat{\mathbf{z}}$ via SQUID magnetometry. The external magnetic field protocol was the same as that used to obtain Figure 2b and d: sufficiently away from the expected $T_M$ (6 K when heating and 300 K when cooling), we applied a magnetic field of 8.5 kOe, then this was lowered to a field $H$, under which we varied the temperature and collected the data. The measured magnetization curves are shown in **Figure S5**, for $H = 8.5, 4.3, 2.1, 0.9$, and 0 kOe. In general terms, the antiferromagnetically coupled Gd and Co magnetizations are equal in magnitude at the compensation point and counteract each other, which explains the overall reduction of $M_z$ around approximately 40 K in the presence of nonzero field.

The thermal hysteresis in the data can be understood from the variations of the free energy across the compensation temperature in a macrospin picture.[7] Well above $T_M$, the Co sublattice magnetization ($\mathbf{M}_{Co}$) is stronger than the Gd sublattice magnetization ($\mathbf{M}_{Gd}$) and, thus, $\mathbf{M}_{Co} \parallel \mathbf{H}$. When the temperature is lowered to a value just below $T_M$ and $|\mathbf{M}_{Gd}| > |\mathbf{M}_{Co}|$, the Zeeman energy favours the alignment of the Gd sublattice with the bias field. However, the sublattice magnetizations do not switch until the Zeeman energy becomes high enough to overcome the energy barrier associated with this reorientation, which occurs at a temperature ($T_r^c$) lower than $T_M$ in a cooling experiment. Consequently, $M_z$ can be negative in a certain temperature range as in Figure S5d. A similar argument can be made for a heating measurement, where the temperature at which the sublattices switch ($T_r^h$) is higher than $T_M$.

The dependence of this phenomenology on the Zeeman energy, which scales with the strength of the magnetic field, accounts for the growing difference between the switching and compensation temperatures for decreasing $H$, as represented in Figure S5. In the limiting case $H = 0$ Oe, the sublattices do not switch and, thus, $M_z$ becomes negative when $|\mathbf{M}_{Gd}|$ surpasses $|\mathbf{M}_{Co}|$.

The qualitative description above neglected the presence of sperimagnetism,[8] domain nucleation, domain wall motion, and the spin-flop transition in GdCo,[7–10] all of which can influence the sublattice magnetization reversal process across $T_M$. The spin-flop transition is particularly noticeable in the minimum of the magnetization curves in Figure S5a. In that experiment, the applied field of 8.5 kOe surpassed the spin-flop field, which significantly decreases around $T_M$.[8,9] Consequently, the sublattice reversal was mediated by a non-collinear alignment of the Co and Gd magnetizations that prevented the vanishing of the $M_z$.[7,10,11]

The magnetization compensation temperature corresponds to the crossing of the heating and cooling branches of the $M_z$ data[7,11] and is virtually independent of the external magnetic field for laboratory-accessible values. In practice, we found small variations of this intersection for different bias fields that we averaged to obtain $T_M = 41.4$ K, denoted by the dashed lines in Figure S5. We estimated an associated uncertainty of $\pm$ 0.3 K by taking the standard deviation of the temperatures at which the crossing occurs.

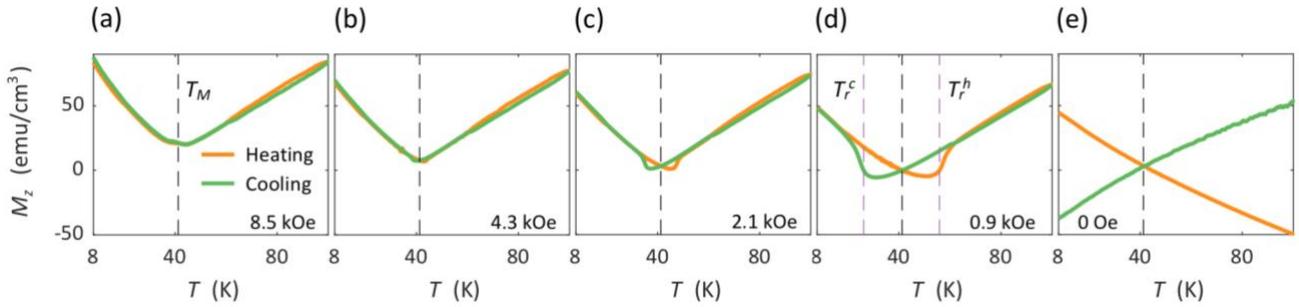

**Figure S5.** a)-e) Static magnetization of the Gd$_{20}$Co$_{80}$(4 nm)/Cu(2 nm)/Pt(3 nm) sample as a function of temperature for bias magnetic fields of 8.5, 4.3, 2.1, 0.9, and 0 kOe, measured by SQUID magnetometry. The black dashed lines in all panels mark the magnetization compensation temperature. The purple dashed lines in d) represent the sublattice reversal temperatures in cooling ($T_r^c$) and heating ($T_r^h$) experiments.

**Supplemental Note S4. THz signal hysteresis loops around magnetization compensation**

We found evidence of the formation of a domain structure in sample 2 (Gd$_{20}$Co$_{80}$(4 nm)/Cu(2 nm)/Pt(3)) close to $T_M$ by measuring the dependence of the THz emission intensity on a magnetic field along $\hat{z}$ across compensation. **Figure S6**a shows that, at sufficiently high (53 K) and low (11 K) temperatures, the obtained $I$ vs $H$ hysteresis loops have typical ferromagnetic-like shapes, and their opposite orientations indicate the complete switching of the magnetic sublattices across compensation. On the other hand, at intermediate temperatures (e.g. 35 K), the loops resemble a superposition of a left-facing loop measured along an easy axis (similarly to that at 53 K) and a right-facing loop measured along a hard axis (similarly to those at 25 K and 11 K). As described next, we interpret the anomalous shape of these loops as a consequence of the coexistence of two types of magnetic domains with opposite alignment of their sublattices, in analogy to the argument presented in Figure 5 of reference[9].

Well above (below) compensation, such as at 53 K (11 K), the Co (Gd) sublattice magnetization aligns with the magnetic field throughout the sample, and the ferromagnetic-like character of the corresponding loops follows from the existence of a single magnetic domain at saturation. The difference in the shape and coercivity between the high and low temperature cases may originate from changes in the magnetic anisotropy, which is known to substantially vary when crossing the compensation line in the composition-temperature phase diagram.[12,13]

The behaviour at intermediate temperatures is determined by the presence of spatial variations in the GdCo composition, a characteristic feature of this material[9,14] that leads to the dependence of $T_M$ on position on the sample. In a cooling experiment, such as that represented in Figure S6, the sublattice magnetizations in regions with above-average Gd concentrations switch at temperatures higher than the mean $T_M$. The volume percentage of this Gd-dominated phase increases progressively as the temperature is decreased at the expense of the Co-dominated phase, until the latter disappears when the temperature is lowered below the minimum $T_M$ in the sample. When the two types of domains coexist, their THz emission interferes, and the resulting $I$ vs $H$ hysteresis loop is a linear combination of the loops corresponding to Co- and Gd-dominated areas, such as those measured at 53 K and 11 K, respectively.

The amplitudes of the two loops in this superposition are proportional to the volume fractions of the two types of domains and, therefore, there is a smooth transition between a simple left-facing (Co-rich) loop to simple (Gd-rich) right-facing loop. We monitored the presence of the Co-dominated contribution by assuming that the coercivity of the loops is mostly due to this type of domains, based on the small coercivity of the low temperature loops (see 25 K data).

The extracted coercive fields as a function of temperature, compared to the corresponding variations of the integrated THz intensity odd with a magnetic field of 8.5 kOe, are displayed in Figure S6b. The coercivity starts to considerably increase below 50 K and peaks at 29 K, which is about 10 K below $T_{\text{THz}} \approx T_{\text{M}}$ where $I(|8.5 \text{ kOe}|)$ is minimum.

At a given temperature, the extracted $H_c$ corresponds to the average over the Co-dominated regions in the sample. Moreover, the coercive field is maximized at $T_{\text{M}}$ (see Figure 4b). Therefore, just below 50 K, which is above the local compensation temperatures in most of the sample, the mean coercivity is small. As the temperature is lowered, increasingly larger regions of the sample become Gd-dominated and approximately stop contributing to the coercivity. At the same time, the remaining Co-dominated areas get closer to compensation, so the average $H_c$ rises. This continues until the Gd sublattice dominates in the whole sample, which we identify with the drop of the mean coercivity just above 25 K.

Hence, we estimate that the two types of domains coexist at temperatures between 25 and 47 K, the upper limit corresponding to the point below which the loops stop exhibiting simple ferromagnetic-like behaviour. Alternatively, this temperature range can be gauged through the width ($W$) of the dip of the $|I|$ vs $T$ plot in Figure S6b (or equivalently in Figure 2a), given that the suppression of the THz signal arises from the cancellation of the emission from the Co- and Gd-dominated phases. Measuring the width from the temperature at which $|I|$ starts to significantly drop to the temperature at which $|I|$ recovers, as illustrated by the dashed lines in Figure S6b, yields the range 27–50 K, in good agreement with the previous estimate. By converting this temperature interval into a GdCo composition difference with the aid of Figure 3a, we estimate that sample 2 has about 1% of variations from its nominal Gd concentration of 20%.

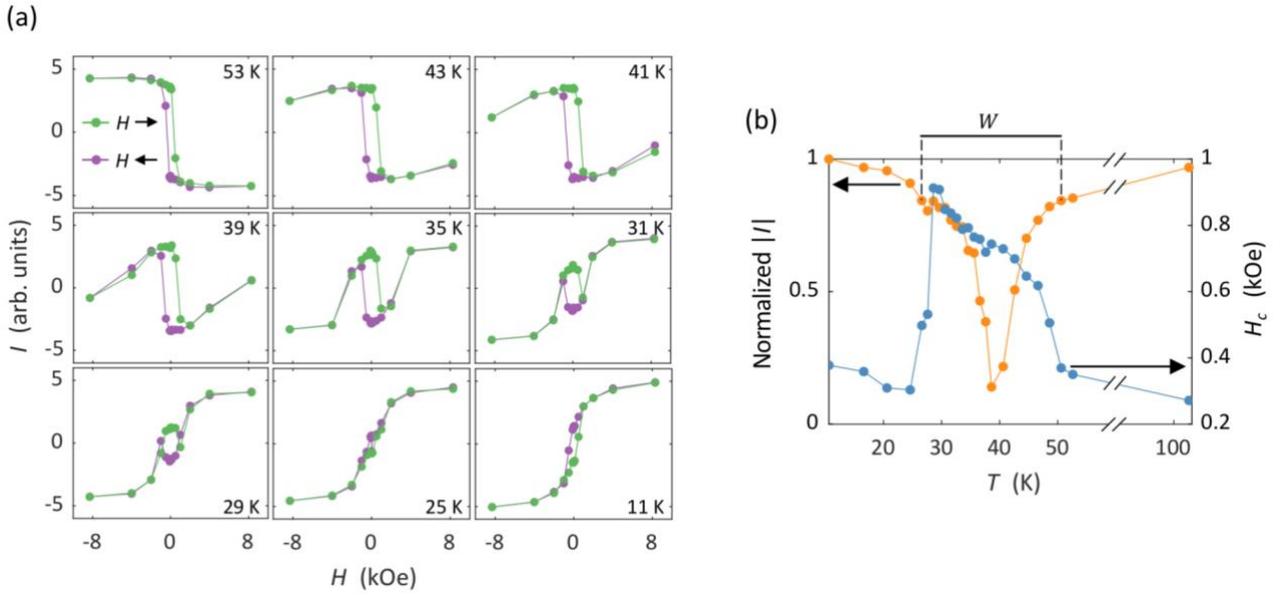

**Figure S6.** Magnetic field dependence of the integrated THz intensity of the Gd$_{20}$Co$_{80}$(4 nm)/Cu(2 nm)/Pt(3)) across magnetization compensation. a) $I$ vs $H$ hysteresis loops at temperatures ranging from 53 K to 11 K. The different colours indicate the branches of the measurements where the magnetic field was increased (green) and decreased (purple) at a fixed temperature. b) Coercive fields and integrated THz intensities extracted from the data set represented in a). The reported intensities were calculated as $[I(8.5\text{ kOe}, T) - I(-8.5\text{ kOe}, T)]/2$ and normalized with respect to the value at 11 K. The vertical dashed line segment on the right indicates the temperature (∼50 K) below which the integrated intensity starts to significantly decrease, and the vertical dashed line segment on the left marks the temperature (∼27 K) at which the intensity recovers to the value it had at 50 K.

**Supplemental Note S5. Polarization of THz signal across magnetization compensation**

The polarization of the THz electric field produced by a STE based on the ISHE reveals the polarization of the spin current responsible for the emission (see Section 2 in the main text). We monitored the evolution of the latter across the magnetization compensation temperature via a THz polarimetry experiment based on the methodology discussed in reference[15]. This consisted in analysing the emitted THz pulses with two wire grid polarizers (WGPs), one at a variable angle and a second aligned with the $\hat{x}$ axis, as depicted in **Figure S7**a. If a given Fourier component of the THz signal produced by the STE is characterized by the Jones vector $\mathbf{u}_i(\nu) = \left(a(\nu)e^{i\delta(\nu)}, b(\nu)\right)$, then Jones calculus can be employed to show that the norm of the vector after transmission through the two WGPs is

$$|\mathbf{u}_f(\nu, \varphi)| = |\cos(\varphi)|[(b(\nu)\sin\varphi + a(\nu)\cos\varphi\cos\delta(\nu))^2 + (a(\nu)\sin\delta(\nu)\cos\varphi)^2]^{1/2}, \text{ (S3)}$$

where $\varphi$ is angle between $\hat{x}$ and the transmission axis of the first WGP. By measuring the THz emission spectrum of sample 2 as a function of $\varphi$ and fitting each frequency component to Equation S3 with $a$, $b$, and $\delta$ as fitting parameters, as exemplified in Figure S7b, we determined the polarization of the generated THz pulses.

We carried out these polarimetry experiments under a bias field $\mathbf{H} = (40\,\hat{x} + 850\,\hat{z})$ Oe. The $\hat{x}$ component was applied to lift the degeneracy of domains with nonzero magnetization along this axis that may form as the temperature is varied. Otherwise, if $\mathbf{H} \cdot \hat{x} = 0$, we would not be able to detect a THz signal originating from this magnetization component as the emission from opposing domains would cancel out.

The obtained polarization states at 24 K and 123 K are presented in Figure S7c and d, where $\theta = \text{atan}(a/b)$. Away from $T_M$ (123 K), the polarization is linear ($\delta = 0$) and orthogonal to $\mathbf{H}$, implying that the net magnetization was parallel to the bias field. Closer to $T_M$ (24 K), the polarization develops a small ellipticity with a major axis approximately at 50° with respect to $\mathbf{H}$, indicating that the polarization of the spin current in the STE forms an angle of about 40° with $\mathbf{H}$.

With the aim of investigating the polarization as a function of temperature in a simpler manner, we exploited the approximate frequency independence in Figure S7c and d and extracted the $\hat{x}$ and $\hat{z}$ polarization amplitudes directly in the time domain. By considering $a$, $b$, and $\delta$ as independent of frequency, it follows from Equation S3 that the $\hat{x}$- and $\hat{z}$-components of the THz signal produced by the STE are $S_x(t) = S_f(t, \varphi = 45°) + S_f(t, \varphi = -45°)$ and $S_z(t) = S_f(t, \varphi = 45°) - S_f(t, \varphi = -45°)$, where $S_f(t, \varphi = \pm 45°)$ is the THz signal with Jones vector $\mathbf{u}_f(\varphi = \pm 45°)$.

Figure S7e shows that, when cooling, the integrated THz intensity of the $\hat{\mathbf{x}}$ component ($I_x$) vanishes at $T_{\text{THz}} \approx 23$ K, similarly to when the applied field is $\mathbf{H} = 8.5\, \hat{\mathbf{z}}$ kOe (Figure 2a), but here the minimum of $I_x$ is about 20 K below $T_M$. In this case, the domain formation that supresses the THz signal does not occur as soon as the regions of the sample with higher Gd concentration cross their compensation point because 850 Oe is weaker than the coercive field in the vicinity of $T_M$. Instead, domains emerge at a lower temperature when the applied field surpasses the local coercivity, enabling the switching of the sublattice magnetizations in these Gd-rich areas.

The integrated THz intensity of the orthogonal polarization component ($I_z$) is negligible away from $T_{\text{THz}}$ (we attribute its small but non-zero intensity to the incomplete attenuation of the polarization perpendicular to the transmission axis of the WGPs). When the temperature is lowered below roughly 90 K, $I_z$ starts to rise, signalling a deviation of the spin polarization from the external magnetic field that grows until ~26 K, where $I_z$ abruptly drops. This value is approximately equal to $T_r^c$ when $\mathbf{H} = 850\, \hat{\mathbf{z}}$ Oe (see Supplemental Material Note S3), the average temperature where the sublattice magnetizations switch. Therefore, the observed canting of the spin polarization can be interpreted as an intermediate step in the reversal of the sublattice magnetization after $T_M$ is crossed.[7]

As mentioned above, the spin rotation can be at least of ~40°, and Figure S7f underscores that the amplitude of the spin projection orthogonal to the magnetic field becomes progressively comparable to the amplitude along the magnetic field until ~26 K. The fact that $I_z$ starts increasing well above $T_M$ may be related to the sperimagnetism of GdCo. In this non-collinear magnetic configuration, the alignment of the sublattice spins is distributed within a cone whose aperture increases close to the magnetization compensation temperature.[8]

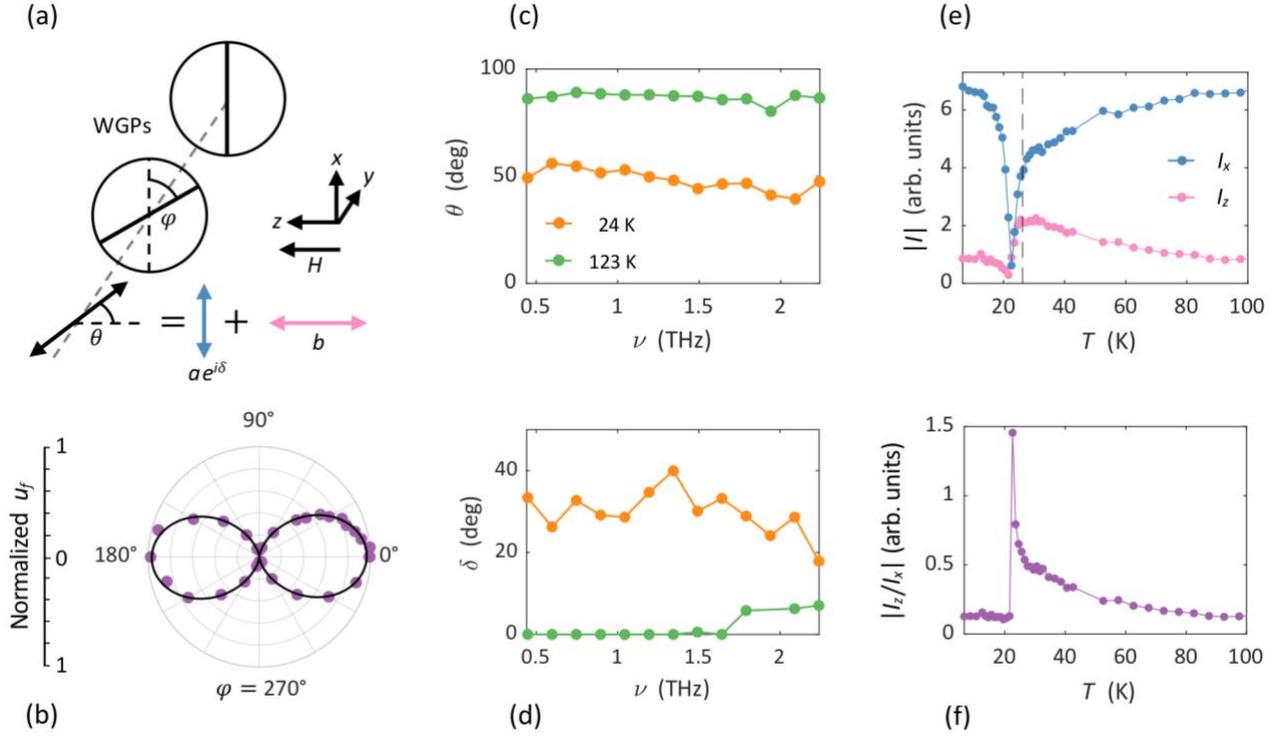

**Figure S7.** Polarization of THz electric field emitted by the $Gd_{20}Co_{80}$(4 nm)/Cu(2 nm)/Pt(3) sample as a function of temperature. a) Schematic of the THz polarimetry experiment used to determine the polarization of the THz pulses. The black double-headed arrow symbolizes the polarization of the THz emission, which forms an angle $\theta$ with respect to the $\hat{z}$ axis in the case where it is linear. b) Example of the polarization analysis done with Jones calculus. The purple circles represent the experimental dependence of THz amplitude on the orientation of the first WGP at 0.9 THz and 123 K. The solid black line is the corresponding fit done with Equation S3. c) and d) Extracted polarization states of the THz signal at 24 K and 123 K. e) Temperature dependence of the integrated THz intensity (defined as in Equation 2 of the main text) of the $\hat{x}$- and $\hat{z}$-components of the THz signal as a function of temperature. f) Temperature dependence of the ratio of the integrated intensity of the $\hat{z}$-component of the THz emission to that of the $\hat{x}$-component.

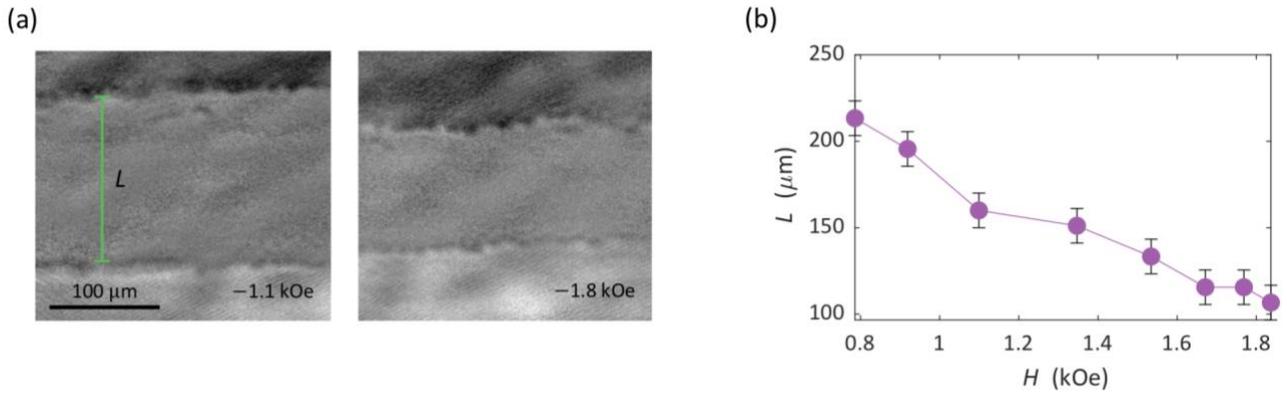

**Figure S8.** Magnetic field dependence of the size of the unsaturated region in the Gd$_{31}$Co$_{69}$(2.9 nm)/Cu(2 nm)/Pt(3 nm) sample close compensation. a) and b) MOKE images of the area of the sample with $T_M$ close to room temperature measured with magnetic fields of $-1.1$ kOe and $-1.8$ kOe along $\hat{\mathbf{z}}$, respectively. The MOKE contrast corresponds to the same type of domains as in Figure 5a in the main text. The micrographs illustrate how the unsaturated area, with width $L$, shrinks as the magnitude of the bias field is increased. c) Width of the unsaturated region as a function of external magnetic field. The error bars of $\pm 10$ μm correspond to the typical difference between the highest and lowest values of $L$ in a given image.

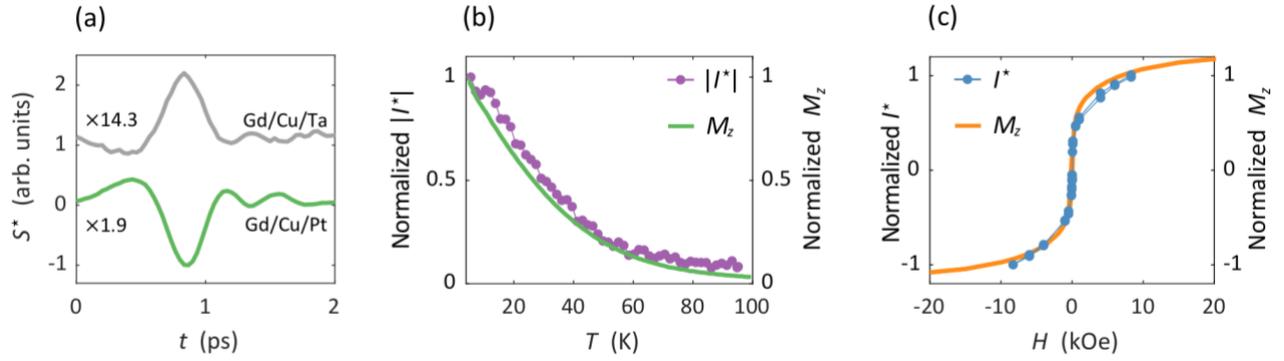

**Figure S9.** THz emission from Gd-based STEs and its connection to the Gd equilibrium magnetization. a) THz pulses from the Gd(4 nm)/Cu(2 nm)/Pt(3 nm) and Gd(4 nm)/Cu(2 nm)/Ta(5 nm) samples measured at 6 K under a bias magnetic field of 8.5 kOe along $\hat{\mathbf{z}}$. The multiplicative factors on the left indicate by how much the waveforms were rescaled to have the same amplitude as the signal of the Co/Cu/Pt sample shown in Figure 6b of the main text. b) and c) Temperature and magnetic field dependence, respectively, of the integrated THz intensity and equilibrium magnetization of the Gd/Cu/Pt sample. The data in b) were obtained with an applied field of 8.5 kOe, and the data in c) were measured at 6 K. The temperature values for the integrated THz intensity data were adjusted in the manner described in Supplemental Material Note S2, which is in principle only valid for the $Gd_{20}Co_{80}$/Cu/Pt sample. The Gd/Cu/Pt sample may absorb and dissipate the laser energy differently, which could be the cause of the small shift between the THz intensity and magnetization data sets in b). The otherwise close correspondence between the $I$ and $M_z$ curves as function of temperature and magnetic field confirms the magnetic origin of the THz emission.